%% file: PoS_ICRC2019_716.tex
\documentclass[a4paper]{PoS}

\usepackage[style=numeric-comp,
			bibstyle=numeric,
			backend=bibtex,
			sorting=none,
			related=false,
			url=false,
			doi=false,
			eprint=false			
			]{biblatex}

\AtEveryBibitem{\clearfield{title}}
\bibliography{ICRC} 

\usepackage{wrapfig}

\usepackage{floatrow}
\newcommand{\ms}{m$^{-2}$\,s$^{-1}$}%
\newcommand\ergs{\ensuremath{\mathrm{erg\,s}^{-1}}}

\newcommand{\dem}{DEM~L241}

\newcommand{\nb}{N\,157B}
\newcommand{\dorc}{30~Dor~C}
\newcommand{\nd}{N\,132D}
\newcommand{\sn}{SN\,1987A}

\newcommand{\pthree}{LMC~P3}

\newcommand{\fermi}{$Fermi$-LAT}
\newcommand{\hess}{H.E.S.S.}

\newcommand{\Msun}{\ensuremath{M_{\odot}}}

\newcommand{\Lacc}{\ensuremath{L_\mathrm{acc}}}

\newcommand{\Mco}{\ensuremath{M_\mathrm{co}}}
\newcommand{\Rco}{\ensuremath{R_\mathrm{co}}}

\newcommand{\MsunYr}{\ensuremath{{\Msun\mathrm{/yr}}}}

\include{plots/includes}

\title{Search for Point-Like TeV Sources in the Large Magellanic Cloud\footnote{The proceedings have been updated on 2020/01/22. In the original version the upper limits on the luminosity and derived values were wrongly calculated.}}

\ShortTitle{Search for Point-Like TeV Sources in the Large Magellanic Cloud}

\author{\speaker{Nukri Komin}\\
        School of Physics, University of the Witwatersrand, Johannesburg, South Africa\\
        E-mail: \email{nukri.komin@wits.ac.za}}
\author{Maria Haupt \\
        DESY, Zeuthen, Germany\\
        E-mail: \email{maria.haupt@desy.de}
        }
\author{for the H.E.S.S. Collaboration \footnote{for collaboration list see PoS(ICRC2019)1177} \thanks{https://www.mpi-hd.mpg.de/hfm/HESS/pages/collaboration/}}

\abstract{The Large Magellanic Cloud (LMC) is an irregular satellite galaxy of the Milky Way, which has been observed extensively in Very-High-Energy (VHE) gamma rays with the H.E.S.S. telescopes since 2004 and reaches now a total observation time of 280 h. The exposure of the LMC is rather inhomogeneous, the region around the Tarantula Nebula having an exposure of up to 220 h while the exposure in the outer parts of the LMC is as low as 5h.

A search for point-like sources was performed on this data set. This search resulted in the detection of the four already known sources (N 157B, N 132D, 30 Dor C and LMC P3) but no further significant emission was revealed. Based on catalogues of pulsars, supernova remnants and high-mass X-ray binaries upper limits on the gamma-ray flux of these objects were derived.

In this talk updated results on the known gamma-ray sources as well as upper limits on the non-detected objects will be presented. It will be shown that for a large part of the LMC the existence of VHE gamma-ray sources with a similar luminosity as the already known sources can be excluded.}

\FullConference{36th International Cosmic Ray Conference -ICRC2019-\\
		July 24th - August 1st, 2019\\
		Madison, WI, U.S.A.}

\begin{document}

\section{Introduction}

The Large Magellanic Cloud is a satellite galaxy of the Milky Way. It has a stellar mass of about 4\% of that of the Milky Way \cite{4}. The LMC is located at a distance of about 50\,kpc \cite{6} and is seen nearly face-on \cite{6}. It harbours a number of potential gamma-ray emitting objects, among those supernova remnants (SNRs) \cite{Maggi2016}, pulsars and pulsar wind nebulae (PWNe) \cite{ATNF, Ridley2013} and High-Mass X-Ray Binaries (HMXB) \cite{Liu2005}.

The LMC is located in the Southern Hemisphere and currently the High Energy Stereoscopic System (\hess) is the only instrument to observe the LMC in TeV gamma rays. \hess has observed the LMC quite extensively since its inauguration in 2004. These observations lead to the discovery of TeV gamma-ray emission from the PWN \nb\ \cite{N157B, Science}, the superbubble \dorc\ \cite{Science}, the SNR \nd\ \cite{Science} and the gamma-ray binary \pthree\ \cite{LMC_P3}. These observations clearly show the presence of TeV gamma-ray emitting sources in the LMC and the capability of \hess\ to detect their emission.

The work presented here searches for further TeV gamma-ray emitting sources in the LMC.

\section{Data Set and Analysis}

\begin{figure}[b]
\begin{minipage}{0.44\textwidth}
\centering
\includegraphics[width=0.8\textwidth]{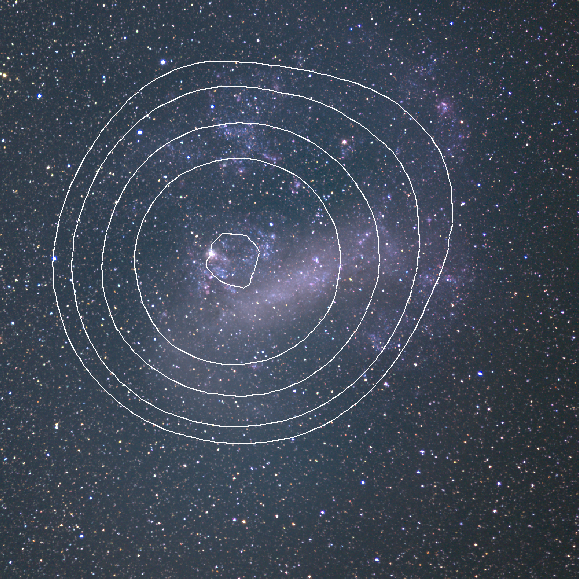} 
\end{minipage}
\begin{minipage}{0.54\textwidth}
\centering
\includegraphics[width=0.8\textwidth]{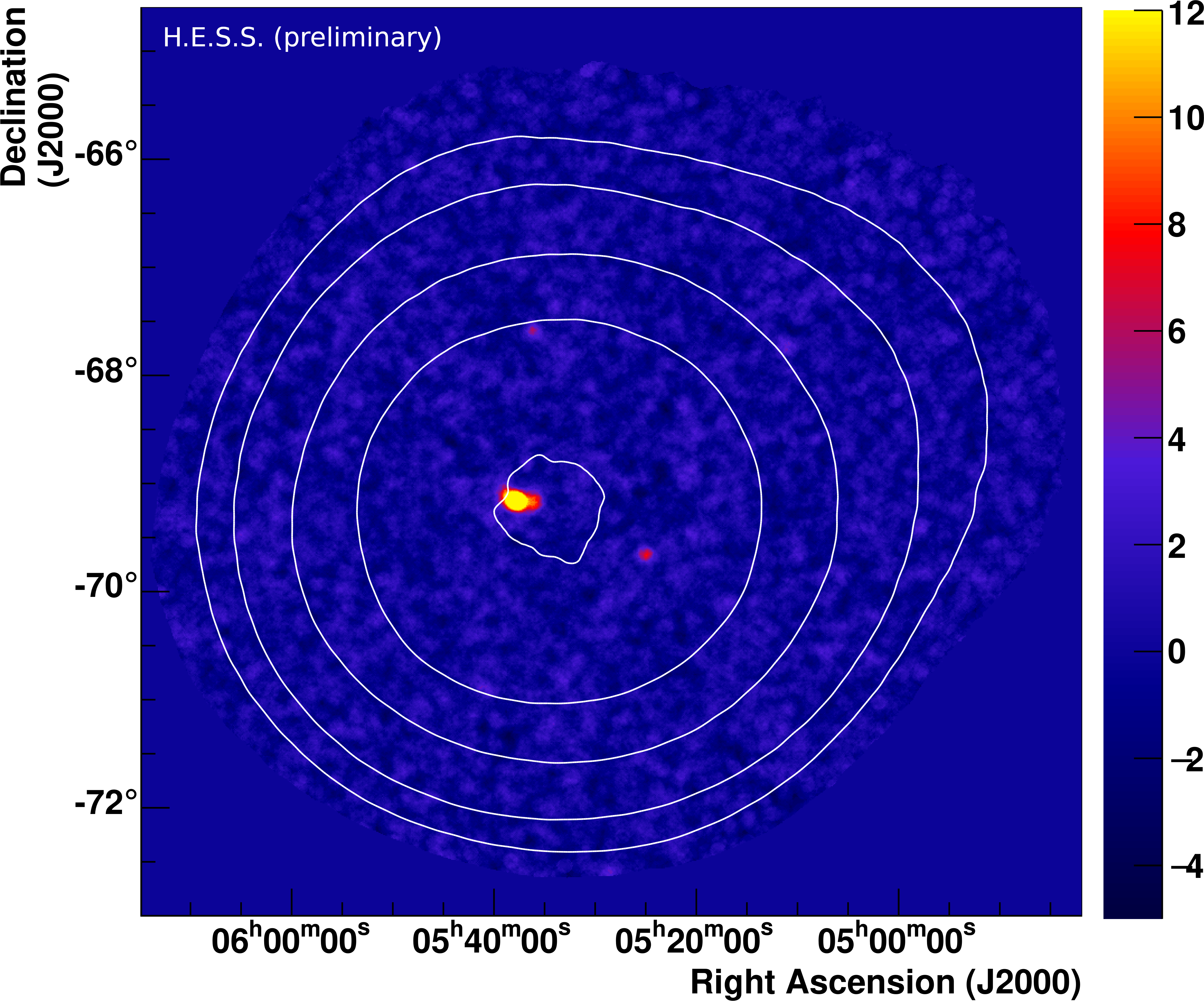} 
\end{minipage}
\caption{\textit{Left panel:} Optical image of the LMC (from \cite{Mellinger}) with overlaid \hess\ exposure contours. The contour lines denote (from outside to inside) 10, 20, 50, 100 and 200\,h effective exposure time. \textit{Right panel:} Significance map of the \hess\ data set with overlaid exposure contours.}
\label{fig:skymap}
\end{figure}

Observations of the LMC with \hess\ are carried out on a nearly yearly basis. While in the beginning the focus was put on the region around the supernova remnant \sn\ and subsequently on the newly detected sources, in the recent years attempts were made to cover as much as possible of the entire LMC. Up to know an exposure of 280\,h has been recorded.

The left panel of Fig.~\ref{fig:skymap} shows the exposure contours overlaid on an optical sky map. It can be seen that the LMC is rather inhomogeneously covered, with about 10\,h of exposure on the western edge and more than 200\,h around the Tarantula nebula and the already detected PWN \nb.

The data were analysed using \textit{Model analysis}
 \cite{Mathieu}, where the camera images are compared with simulations using a log-likelihood minimisation.
 In addition to the standard cuts an optional cut on the reconstructed direction error was applied in order to increase the angular resolution and further decrease the background. A similar cut was used in previous publications \cite{Science,LMC_P3}. The background was estimated from rings around each sky position to generate the gamma-ray image \cite{BGmodels}. The energy threshold for this data set is 714\,GeV.

The right panel of Fig.~\ref{fig:skymap} shows the significance map for this data set. It is the first time that a complete view of the LMC in TeV gamma rays is presented.

\section{The Pulsar Wind Nebula N\,157B}

\begin{wrapfigure}{r}{0.5\textwidth}
\includegraphics[width=0.9\textwidth]{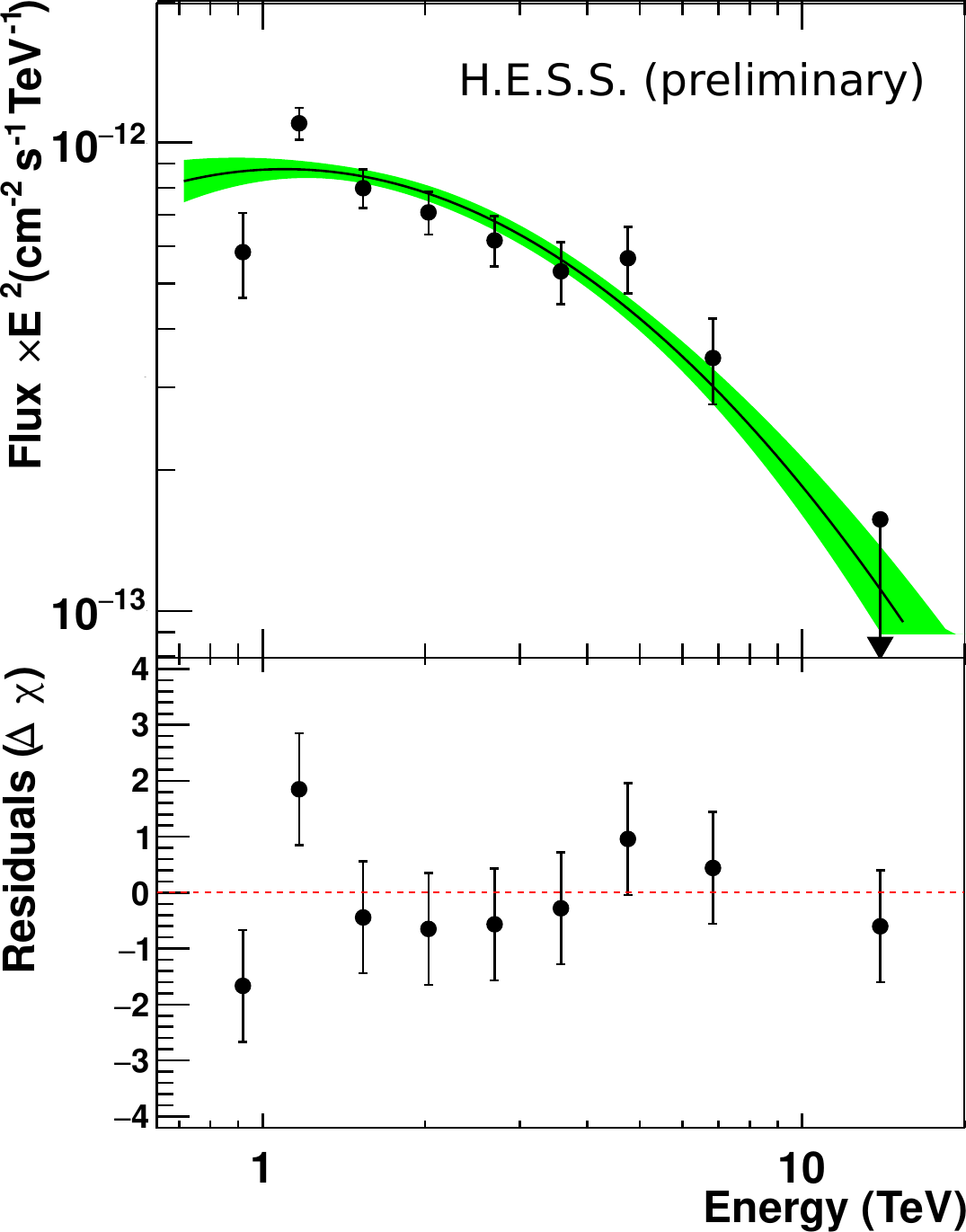} 
\caption{Gamma-ray energy spectrum of the PWN \nb.}
\label{fig:N157B_spectrum}
\end{wrapfigure}

In the significance map in Fig.~\ref{fig:skymap} the emission from the PWN \nb\ can be clearly seen. With a flux of about 2\% of the flux of the Crab Nebula \nb\ is by far the brightest VHE gamma ray source in the LMC. It is detected in yearly subsets of the deep data set of the LMC and can therefore be used as a standard candle for the LMC.

For the study of this particular object the additional high-resolution cut was not applied, in order to obtain more excess counts. In the current data set 1152 excess events with a statistical significance of $28.6\sigma$ are detected from \nb.
Figure~\ref{fig:N157B_spectrum} shows the gamma-ray energy spectrum of this object. The spectrum can be fit with a curved power law
\begin{equation}
\frac{dN}{dE} = \Phi_0
\left( \frac{E}{1.5\mathrm{TeV}} \right)
^{- \alpha - \beta \times \ln{\left( \frac{E}{1.5\mathrm{TeV}} \right)} }
,
\end{equation}
with $\Phi_0 = \left( 3.6 \pm 0.1 \right) \times 10^{-9}\mathrm{m}^{-2}\mathrm{s}^{-1}\mathrm{TeV}^{-1}$, $\alpha = \left( 2.23 \pm 0.08 \right)$ and $\beta = \left( 0.32 \pm 0.07 \right)$ .
This three-parameter model is preferred by $4.4\sigma$ over a simple power law. The gamma-ray luminosity in the $1 - 10$\,TeV energy range for a distance of 50\,kpc is $(6.5 \pm 0.3) \times 10^{35}$\,erg/s and matches within errors the previously reported value \cite{Science}.

%
%

\section{Upper Limits on Gamma-Ray Fluxes from Selected Sources}

\begin{figure}[t]
\begin{minipage}{0.49\textwidth}
\includegraphics[width=\textwidth]{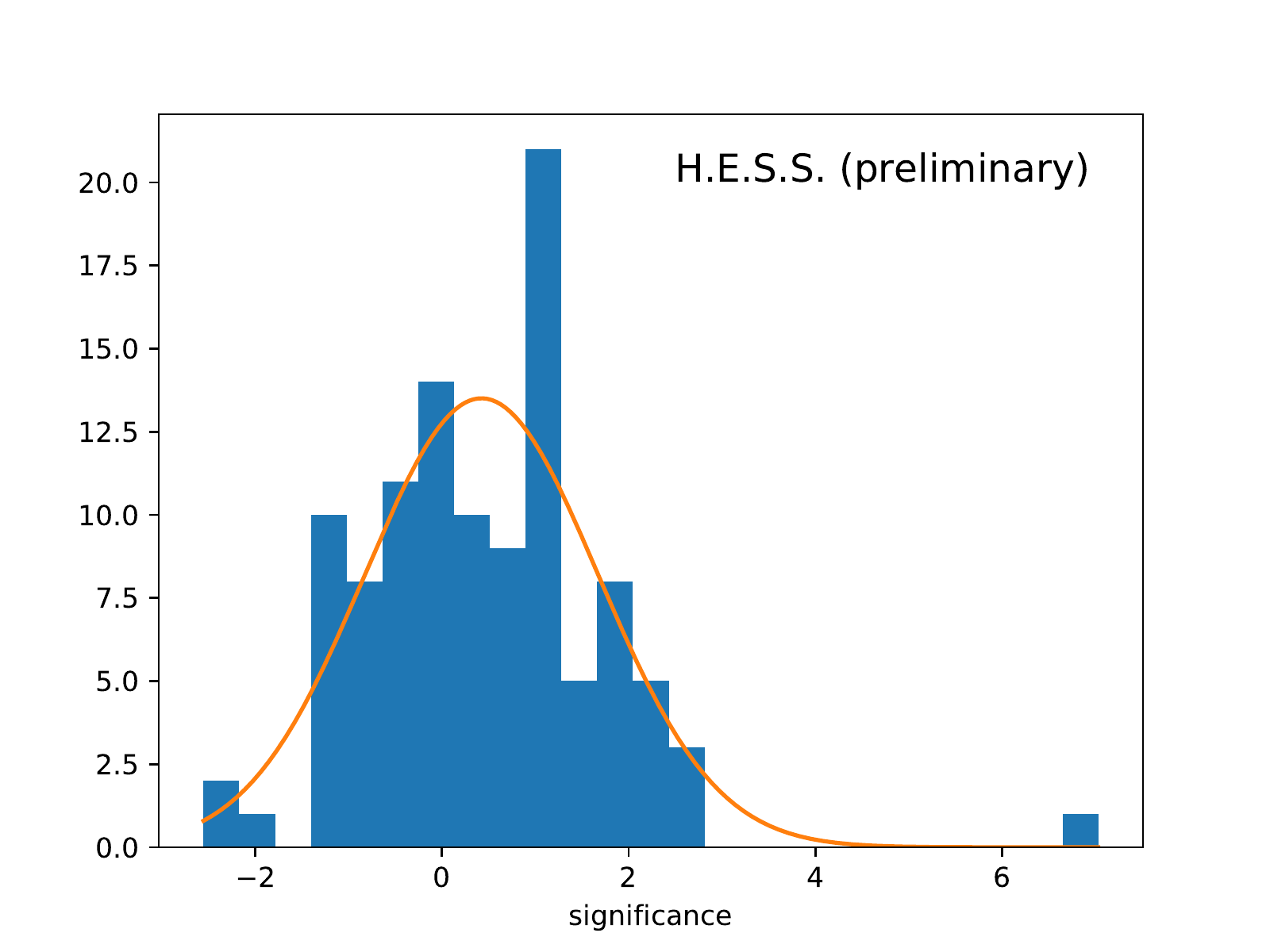} 
\end{minipage}
\begin{minipage}{0.49\textwidth}
\includegraphics[width=\textwidth]{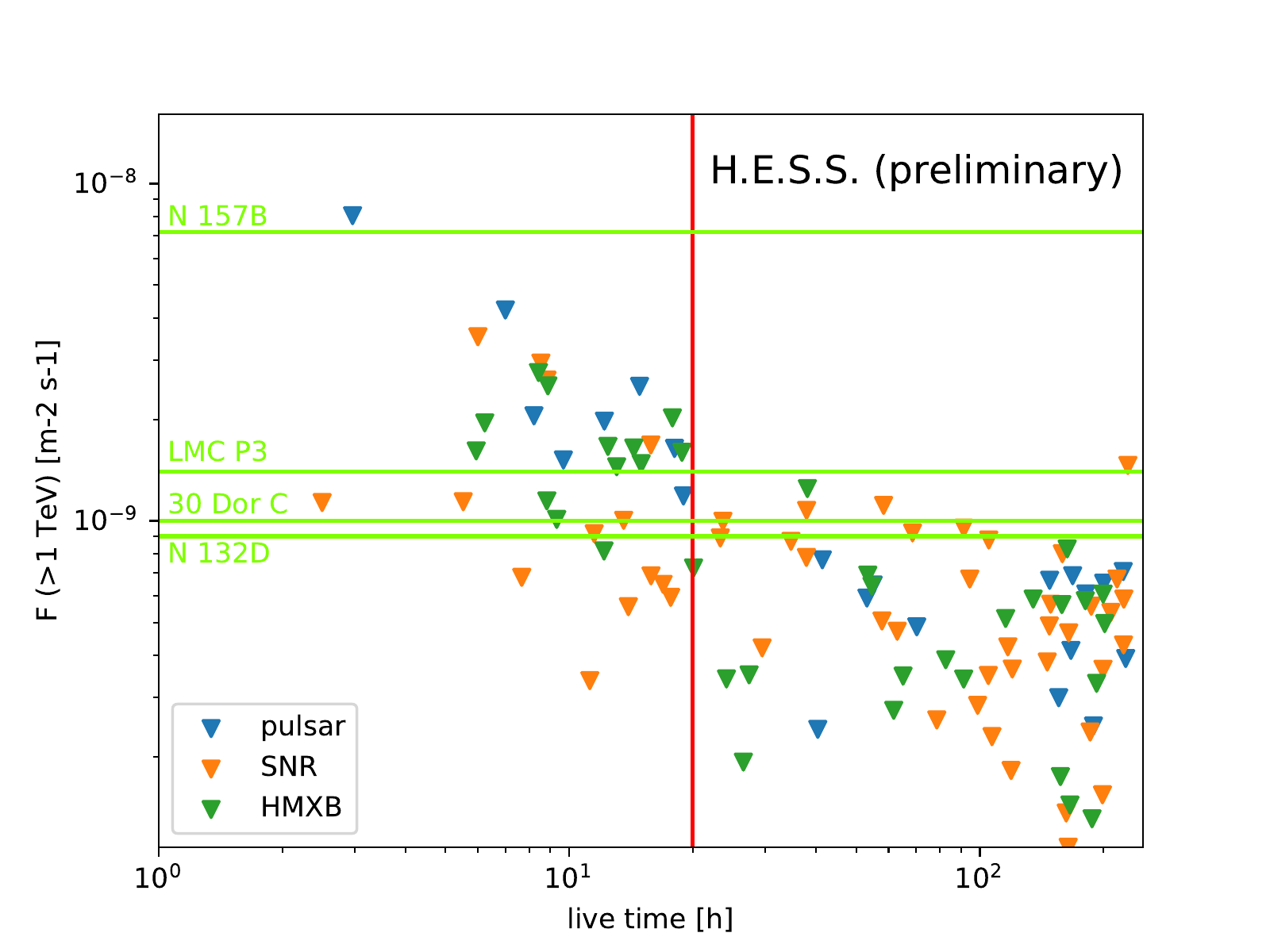}
\end{minipage}
\caption{\textit{Left panel:} Significance distribution of the all objects (excluding the detected sources). \textit{Right panel:} Upper limits on the gamma-ray flux from pulsars, SNRs and HMXB depending on the exposure time of the observations. The red line indicates the limit of 20\,h of exposure time. The green line indicate the flux levels of the already detected sources. These limits are taken from \cite{Science} for \nb, \dorc and \nd\ and from \cite{LMC_P3} for \pthree.}
\label{fig:UL_vs_T}
\end{figure}

The data set is searched for emission from pulsars, SNRs and HMXBs. For each of the objects found in the catalogues a dedicated analysis was performed. No significant emission was found from any of the objects. The left panel of Fig.~\ref{fig:UL_vs_T} shows the significance distribution which can be fit by a Gaussian with a mean of 0.4 and a standard deviation of 1.2. All known sources have been removed from this plot. The only outlying object is SNR~B0536$-$6914 with a statistical significance of $6.6\,\sigma$. A closer inspection of the object shows that this is spill-over emission from \nb\ and \dorc\ and not genuine emission from this object itself.

Upper limits on the gamma-ray photon flux between 1 and 10\,TeV for a confidence level of 95\% and an assumed spectral index of -2.3 are computed for all objects.
The right panel of Fig.~\ref{fig:UL_vs_T} shows the derived upper limits depending on the corresponding exposure time for each object. Green lines denote flux levels of detected sources. Further sources with flux levels similar to \nb\ can be excluded. Where the exposure time exceeds 20\,h it can also be excluded that sources exhibit a gamma-ray flux at the level of \dorc, \nd\ or \pthree.

It can be seen that for more than 20\,h exposure only marginal improvement of the upper limits can be achieved with increasing exposure time. Therefore, an increase of exposure time beyond 20\,h will not increase the detection capability of sources in the LMC. On the other side, the detection of new sources is still possible by increasing the exposure time to 20\,h everywhere in the LMC. In the following only upper limits on objects with a minimum exposure of at least 20\,h are reported.

\subsection{Pulsars and Pulsar Wind Nebulae}

The loss of rotational energy of pulsars can power PWNe. These nebulae are the most abundant emitters of galactic TeV emission (see \cite{HGPS,PWN}). Further on, pulsed gamma-ray emission in the 100\,GeV to TeV range has been detected from a few pulsars themselves: the Crab pulsar \cite{Crab1, Crab2}, the Vela pulsar \cite{Vela} and most recently from Geminga. These findings motivate to search for TeV gamma-ray emission near pulsars in the LMC.

The \hess\ LMC data set has been analysed at source positions corresponding to the pulsars found in the ATNF pulsar data base \cite{ATNF} and in \cite{Ridley2013}. Apart from \nb\ (see above) no significant emission from any other object was detected. The significances and upper limits on the photon flux are reported in Table~\ref{tab:pulsars}. The photon flux limits between 1 and 10\,TeV are below $10^{-9}$\ms, well below the photon flux from \nb\ of $7.5 \times 10^{-9}$\ms. The upper limit on the gamma-ray luminosity, $L$, can be related to the spin-down luminosity, $\dot E$, of the pulsar, resulting in an upper limit on the pulsar efficiency. These efficiencies are typically of the order of one percent \cite{Emma}, it is 0.08\% for \nb\ \cite{N157B}. The upper limits on the efficiencies in Table~\ref{tab:pulsars} are not constraining for most of the pulsars in the LMC.  The most remarkable object in this list is PSR J0540$-$6919, the pulsar with the third highest spin-down luminosity and for which gamma-ray pulsations have been detected by \fermi\ \cite{FermiPulsar}. If this pulsar is a TeV gamma-ray emitter, its efficiency is extremely low. All other pulsars are extremely old, several hundred kiloyears, so gamma-ray emission is not necessarily expected from these objects.

\input{plots/psr_table.tex}

\input{plots/floatrow}

\subsection{Supernova Remnants}

SNRs are the prime candidates for the acceleration sites of the galactic Cosmic Rays. TeV emission has been detected from a number of SNRs in the Milky Way and from \nd\ in the LMC. This motivates the search for gamma-ray emission from SNRs in the LMC. Here the catalogue of SNRs in the LMC provided by \cite{Maggi2016} is used. Three of the SNRs show significant emission: The already known \nd\ and \dem. The emission at the position of the latter is variable and periodic and can be attributed to the binary \pthree\ inside the SNR \cite{LMC_P3}. The emission at the position of SNR~B0536$-$6914 is due to spill-over from neighbouring sources. These three objects are removed from the list. The analysis of SNR \sn\ does suffer from spill-over as well, but to a much lesser degree than SNR~B0536$-$6914. The upper limits on \sn\ are therefore very conservative.

Table~\ref{tab:snr} lists the upper limits on the photon flux and luminosity of the remaining objects.
The upper limit on the photon fluxes are of the order of the flux of \nd. It can therefore be excluded that any of these SNRs emits TeV gamma-ray emission similar to \nd.
The corresponding total energy in protons is calculated for an assumed ambient density of $n = 1\mathrm{cm}^{-3}$ with a cooling time of
\begin{equation}
\tau_{pp} = 4\times10^{15}\left( \frac{n}{1\,\mathrm{cm}^{-3}} \right)^{-1}\,\mathrm{s}.
\end{equation}
It can be seen from Table~\ref{tab:snr} that the total energy in protons roughly corresponds to the canonical 10\% of the explosion energy of $10^{51}\mathrm{erg}$. It should be noted that the gamma-ray upper limits are for the 1--to--10\,TeV energy range and therefore probing protons above $\sim10\,\mathrm{TeV}$ only. Any lower energy protons are not considered in the limits presented here. Further on, a higher ambient target density would lower the required total energy in protons.

\subsection{High-Mass X-Ray Binaries}

Gamma-ray binaries are a growing class of HMXB systems exhibiting strong emission in the gamma-ray energy regime. The latest discovery is the binary system \pthree\ in the LMC \cite[and references therein]{LMC_P3}. This motivates a search for TeV gamma-ray emission from other HMXBs in the LMC. The positions of the HMXBs in \cite{Liu2005} are analysed. No significant emission from any of these objects is found and upper limits on the photon flux are shown in Table~\ref{tab:hmxb}. The upper limit on the photon fluxes are of the order of the flux of \pthree. It can therefore be excluded that any of these binaries emits TeV gamma-ray emission similar to \pthree. However, this study considers only the average flux over the entire orbit. \pthree\ emits only during 20\% or less of its orbit. There for a search for pulsed emission from the binary systems still has discovery potential.

The upper limit on the luminosity of the HMXBs can be compared with the accretion luminosity of stellar wind from the companion star onto a hypothetical pulsar,
\begin{equation}
\Lacc = 
\left( \frac{\dot{M} }{ 10^{-10} \MsunYr} \right) 
\left( \frac{\Mco}{ 1.4 \Msun } \right) 
\left( \frac{10 \mathrm{km} }{\Rco} \right) \times 
  1.2 \times 10^{36} \,\ergs,
\end{equation}
where \Mco\ and \Rco\ are the mass and radius of the compact object \cite{Accretion}. $\dot{M}$ is the mass-accretion rate which is of the order of $10^{-10} \MsunYr$ in a typical close binary system with accretion of the stellar wind. The upper limits on the luminosities of the HMXBs corresponds to a few percent of this accretion efficiency. If accretion is taking place in these systems then the conversion of accretion luminosity into TeV gamma-rays is lower then a few percent. However, a much higher accretion rate during only a part of the orbit cannot be ruled out.

\section{Conclusion}

The \hess\ Collaboration is expanding the observations of the LMC in order to achieve a full survey of the galaxy. It is shown that a minimum exposure of 20\,h is sufficient to test for emission at the level of the already detected sources. The search for emission from pulsars, SNRs and HMXB did not reveal any new sources. Upper limits on these objects are presented where a minimum exposure of 20\,h of exposure was achieved. These upper limits probe to some extent the physical parameter space of gamma-ray production in these objects.

The increased exposure on the PWN \nb\ shows a clear curvature of the spectrum. More detailed modelling of the emission including the measurements at lower energies carried by \fermi\ will be performed.

The work presented here defines the baseline for the continued observations with \hess\ as well as future observations with the Cherenkov Telescope Array (CTA). With the \hess\ telescopes only emission at flux levels of the already detected sources can be measured.  CTA will be able to probe for much fainter objects \cite{CTA_LMC}.


\section*{\small Acknowledgements}
\small

The support of the Namibian authorities and of the University of Namibia in facilitating the construction and operation of H.E.S.S. is gratefully acknowledged, as is the support by the German Ministry for Education and Research (BMBF), the Max Planck Society, the German Research Foundation (DFG), the Helmholtz Association, the Alexander von Humboldt Foundation, the French Ministry of Higher Education, Research and Innovation, the Centre National de la Recherche Scientifique (CNRS/IN2P3 and CNRS/INSU), the Commissariat à l'énergie atomique et aux énergies alternatives (CEA), the U.K. Science and Technology Facilities Council (STFC), the Knut and Alice Wallenberg Foundation, the National Science Centre, Poland grant no. 2016/22/M/ST9/00382, the South African Department of Science and Technology and National Research Foundation, the University of Namibia, the National Commission on Research, Science \& Technology of Namibia (NCRST), the Austrian Federal Ministry of Education, Science and Research and the Austrian Science Fund (FWF), the Australian Research Council (ARC), the Japan Society for the Promotion of Science and by the University of Amsterdam.

We appreciate the excellent work of the technical support staff in Berlin, Zeuthen, Heidelberg, Palaiseau, Paris, Saclay, Tübingen and in Namibia in the construction and operation of the equipment. This work benefitted from services provided by the H.E.S.S. Virtual Organisation, supported by the national resource providers of the EGI Federation. 



\printbibliography

\end{document}

%% file: plots/includes.tex
\def\utw{\smash{\rlap{\lower5pt\hbox{$\sim$}}}}
\def\udtw{\smash{\rlap{\lower6pt\hbox{$\approx$}}}}

\def\sun{{\lower-2pt\hbox{$_\odot$}}}

\def\g{$\gamma$}

%% file: plots/psr_table.tex
\begin{table}
\small\begin{center}
\caption{Results from the analysis of pulsars in the LMC. The upper limit on the luminosity is calculated for an assumed distance of 50\,kpc. The spin-down luminosities are taken from ATNF \cite{ATNF}, a dash indicates no available value. \label{tab:pulsars}}
\begin{tabular}{cccccc}
\hline \hline
name & Sig & UL & L & Edot & PSR eff \\
 &  & $\mathrm{s^{-1}\,m^{-2}}$ & $\mathrm{erg\,s^{-1}}$ & $\mathrm{erg\,s^{-1}}$ & $\mathrm{}$ \\
\hline\hline
J0519-6932 & 1.2 & 6.9e-10 & 7.5e+34 & 1.5e+33 & 5.0e+01 \\
J0521-68 & 1.1 & 4.1e-10 & 4.5e+34 & - & - \\
J0522-6847 & 1.1 & 6.1e-10 & 6.6e+34 & 2.3e+33 & 2.9e+01 \\
J0529-6652 & -0.1 & 5.9e-10 & 6.4e+34 & 6.6e+32 & 9.8e+01 \\
J0532-6639 & -1.2 & 2.4e-10 & 2.6e+34 & 8.0e+32 & 3.3e+01 \\
J0532-69 & 0.2 & 6.7e-10 & 7.3e+34 & - & - \\
J0534-6703 & -0.3 & 4.8e-10 & 5.3e+34 & 2.8e+33 & 1.9e+01 \\
J0535-66 & 0.9 & 6.5e-10 & 7.0e+34 & - & - \\
J0535-6935 & -0.1 & 3.9e-10 & 4.3e+34 & 5.6e+34 & 7.6e-01 \\
J0537-69 & 1.9 & 7.1e-10 & 7.7e+34 & - & - \\
J0540-6919 & 2.5 & 6.5e-10 & 7.1e+34 & 1.5e+38 & 4.9e-04 \\
J0542-68 & 1.0 & 3.0e-10 & 3.3e+34 & - & - \\
J0543-6851 & -0.7 & 2.5e-10 & 2.7e+34 & 4.4e+32 & 6.2e+01 \\
J0555-7056 & -0.2 & 7.7e-10 & 8.4e+34 & 4.1e+32 & 2.0e+02 \\
\hline\hline
\end{tabular}
\end{center}
\end{table}

%% file: plots/floatrow.tex
\begin{table}
\tiny
  \begin{floatrow}
    \ttabbox{
      \caption{Results from the analysis of SNRs in the LMC. The upper limit on the luminosity is calculated for an assumed distance of 50\,kpc. The upper limit on the total energy is calculated with a cooling time of $\tau_{pp} = 4\times10^{15}\,\mathrm{s}$ assuming an ambient target density of $1\,\mathrm{cm}^{-3}$.}
        \label{tab:snr}
      }{

\begin{tabular}{ccccc}
\hline \hline
name & Sig & UL & L & Wtot \\
 &  & $\mathrm{s^{-1}\,m^{-2}}$ & $\mathrm{erg\,s^{-1}}$ & $\mathrm{erg}$ \\
\hline\hline
1RXSJ053353.6-7204 & 1.3 & 1.0e-09 & 1.1e+35 & 4.3e+50 \\
B0509-67.5 & 1.0 & 1.1e-09 & 1.2e+35 & 4.7e+50 \\
B0519-690 & -0.2 & 4.7e-10 & 5.1e+34 & 2.0e+50 \\
B0520-694 & 1.4 & 8.0e-10 & 8.7e+34 & 3.5e+50 \\
B0528-6716 & -1.3 & 2.6e-10 & 2.8e+34 & 1.1e+50 \\
B0528-692 & -2.6 & 1.5e-10 & 1.7e+34 & 6.7e+49 \\
B0532-675 & -0.5 & 1.8e-10 & 2.0e+34 & 7.9e+49 \\
B0534-699 & -1.1 & 5.3e-10 & 5.8e+34 & 2.3e+50 \\
B0536-6914 & 7.0 & 1.5e-09 & 1.6e+35 & 6.4e+50 \\
B0540-693 & 2.6 & 6.7e-10 & 7.3e+34 & 2.9e+50 \\
B0548-704 & -0.0 & 3.6e-10 & 4.0e+34 & 1.6e+50 \\
DEM L109 & -0.7 & 3.5e-10 & 3.8e+34 & 1.5e+50 \\
DEM L175a & -0.0 & 8.9e-10 & 9.7e+34 & 3.9e+50 \\
DEM L205 & 0.5 & 2.8e-10 & 3.1e+34 & 1.2e+50 \\
DEM L214 & 0.0 & 5.0e-10 & 5.5e+34 & 2.2e+50 \\
DEM L218 & -1.3 & 2.4e-10 & 2.6e+34 & 1.0e+50 \\
DEM L238 & -0.5 & 3.8e-10 & 4.2e+34 & 1.7e+50 \\
DEM L249 & 1.1 & 5.7e-10 & 6.2e+34 & 2.5e+50 \\
DEM L256 & -0.3 & 4.2e-10 & 4.6e+34 & 1.8e+50 \\
DEM L299 & 2.7 & 5.6e-10 & 6.1e+34 & 2.4e+50 \\
DEM L316A & -1.4 & 1.4e-10 & 1.5e+34 & 5.9e+49 \\
DEM L316B & -2.0 & 1.1e-10 & 1.2e+34 & 4.7e+49 \\
DEM L71 & 0.7 & 8.7e-10 & 9.5e+34 & 3.8e+50 \\
Honeycomb & 0.2 & 4.3e-10 & 4.7e+34 & 1.9e+50 \\
J0508-6830 & -0.1 & 4.7e-10 & 5.1e+34 & 2.1e+50 \\
J0511-6759 & 2.1 & 1.1e-09 & 1.2e+35 & 4.8e+50 \\
J0514-6840 & 2.0 & 9.5e-10 & 1.0e+35 & 4.2e+50 \\
J0517-6759 & 2.2 & 6.7e-10 & 7.3e+34 & 2.9e+50 \\
J0550-6823 & 0.1 & 2.3e-10 & 2.5e+34 & 1.0e+50 \\
N103B & 1.8 & 9.2e-10 & 1.0e+35 & 4.0e+50 \\
N120 & -0.5 & 4.9e-10 & 5.3e+34 & 2.1e+50 \\
N159 & -0.6 & 3.6e-10 & 4.0e+34 & 1.6e+50 \\
N206 & 1.4 & 8.8e-10 & 9.6e+34 & 3.8e+50 \\
N23 & 0.9 & 7.8e-10 & 8.5e+34 & 3.4e+50 \\
N44 & 1.1 & 4.2e-10 & 4.6e+34 & 1.8e+50 \\
SNR1987A & 1.5 & 5.9e-10 & 6.4e+34 & 2.6e+50 \\
\hline\hline
\end{tabular}
        
      }

    \ttabbox{
      \caption{Upper limits on high-mass X-ray binaries in the LMC.}
        \label{tab:hmxb}
      }{

\begin{tabular}{cccc}
\hline \hline
name & Sig & UL & L \\
 &  & $\mathrm{s^{-1}\,m^{-2}}$ & $\mathrm{erg\,s^{-1}}$ \\
\hline\hline
1A 0535-668 & 1.1 & 6.4e-10 & 7.0e+34 \\
1SAX J0544.1-7100 & 0.2 & 3.9e-10 & 4.2e+34 \\
2A 0532-664 & -0.2 & 1.9e-10 & 2.1e+34 \\
3A 0540-697 & 0.7 & 6.1e-10 & 6.6e+34 \\
H 0544-665 & -0.3 & 3.5e-10 & 3.8e+34 \\
RX J0456.9-6824 & 0.6 & 7.3e-10 & 7.9e+34 \\
RX J0507.6-6847 & -1.3 & 2.7e-10 & 3.0e+34 \\
RX J0512.6-6717 & 1.9 & 1.2e-09 & 1.4e+35 \\
RX J0516.0-6916 & 0.4 & 5.9e-10 & 6.4e+34 \\
RX J0520.5-6932 & 1.9 & 8.3e-10 & 9.0e+34 \\
RX J0523.2-7004 & 0.2 & 5.6e-10 & 6.2e+34 \\
RX J0524.2-6620 & -2.2 & 3.4e-10 & 3.7e+34 \\
RX J0527.1-7005 & 1.0 & 5.8e-10 & 6.3e+34 \\
RX J0529.4-6952 & 0.4 & 5.0e-10 & 5.4e+34 \\
RX J0532.3-7107 & -0.5 & 3.4e-10 & 3.7e+34 \\
RX J0535.0-6700 & -0.7 & 3.5e-10 & 3.8e+34 \\
RX J0535.6-6651 & 1.3 & 6.9e-10 & 7.5e+34 \\
RX J0541.4-6936 & -0.9 & 3.3e-10 & 3.6e+34 \\
RX J0541.5-6833 & -1.4 & 1.3e-10 & 1.4e+34 \\
RX J0546.8-6851 & 0.1 & 1.7e-10 & 1.9e+34 \\
XMMU J053115.4-705350 & 0.8 & 5.1e-10 & 5.6e+34 \\
XMMU J054134.7-682550 & -0.7 & 1.4e-10 & 1.6e+34 \\
\hline\hline
\end{tabular}

    }

  \end{floatrow}
  \end{table}